\documentclass[preprintnumbers, floatfix,letterpaper,aps,prd,epsfig,nofootinbib, twocolumn]{revtex4-1}
\usepackage{bm,graphicx,dcolumn,epstopdf,epsf, latexsym,mathbbol, amssymb,amsmath, color, slashed, mathrsfs,mathcomp,simplewick}
\usepackage[center]{subfigure}
\usepackage{multirow}
\usepackage{makecell}
\usepackage{ytableau}
\usepackage{booktabs}
\usepackage[colorlinks,linkcolor=blue,citecolor=blue,urlcolor=blue]{hyperref}

\begin{document}
\allowdisplaybreaks
 \newcommand{\bq}{\begin{equation}}
 \newcommand{\eq}{\end{equation}}
 \newcommand{\bqn}{\begin{eqnarray}}
 \newcommand{\eqn}{\end{eqnarray}}
 \newcommand{\nb}{\nonumber}
 \newcommand{\lb}{\label}
 \newcommand{\f}{\frac}
 \newcommand{\p}{\partial}
\newcommand{\PRL}{Phys. Rev. Lett.}
\newcommand{\PLB}{Phys. Lett. B}
\newcommand{\PRD}{Phys. Rev. D}
\newcommand{\CQG}{Class. Quantum Grav.}
\newcommand{\JCAP}{J. Cosmol. Astropart. Phys.}
\newcommand{\JHEP}{J. High. Energy. Phys.}
\newcommand{\bea}{\begin{eqnarray}}
\newcommand{\ena}{\end{eqnarray}}
\newcommand{\beqa}{\begin{eqnarray}}
\newcommand{\eeqa}{\end{eqnarray}}
\newcommand{\red}{\textcolor{black}}

\newlength\scratchlength
\newcommand\s[2]{
  \settoheight\scratchlength{\mathstrut}%
  \scratchlength=\number\numexpr\number#1-1\relax\scratchlength
  \lower.5\scratchlength\hbox{\scalebox{1}[#1]{$#2$}}%
}


\title{Signatures of Quantum-Corrected Black Holes in Gravitational Waves from Periodic Orbits}

\author{Fazlay Ahmed${}^{a, b}$}
\email{fazleyahmad12@gmail.com}

\author{Qiang Wu${}^{a, b}$}
\email{wuq@zjut.edu.cn}

\author{Sushant~G.~Ghosh${}^{c, d}$}
\email{sghosh2@jmi.ac.in}

\author{Tao Zhu${}^{a, b}$ }
\email{Corresponding author: zhut05@zjut.edu.cn}

\affiliation{${}^{a}$Institute for Theoretical Physics \& Cosmology, Zhejiang University of Technology, Hangzhou, 310023, China\\
${}^{b}$ United Center for Gravitational Wave Physics (UCGWP),  Zhejiang University of Technology, Hangzhou, 310023, China\\
${}^{c}$Centre for Theoretical Physics, Jamia Millia Islamia, New Delhi 110025, India\\
${}^{d}$ Astrophysics and Cosmology Research Unit, School of Mathematics, Statistics and Computer Science, University of KwaZulu-Natal, Private Bag 54001, Durban 4000, South Africa}

\date{\today}

\begin{abstract}

Focusing on a Schwarzschild spacetime modified by loop quantum gravity (LQG) holonomy corrections, this work explores the gravitational radiation produced by test masses in bound periodic configurations. The underlying geometry incorporates a quantum-correction parameter $q_c$ that alters the radial metric scaling while maintaining both asymptotic flatness and the classical horizon radius.
The bound geodesics are systematically classified using the zoom--whirl representation labelled by three integers $(z,w,v)$. Gravitational waveforms are computed within a numerical framework that combines exact geodesic motion with the quadrupole approximation, which is suitable for extreme mass ratio inspirals. We demonstrate that the quantum corrections lead to distinct phase shifts, amplitude variations, and modifications to the harmonic structure of the waveforms, with increasingly complex features for orbits with larger zoom numbers. The corresponding frequency spectra and characteristic strain peak, which fall within the millihertz band, are within the sensitivity ranges of space-based detectors such as LISA, Taiji, and TianQin. For specific orbital configurations and values of the quantum-correction parameter, the characteristic strain exceeds the projected detector noise, indicating potential observability. Our results demonstrate that gravitational waves from periodic orbits provide a sensitive probe of quantum-corrected black hole spacetimes in the strong-field regime.

\end{abstract}


\maketitle

\section{Introduction}
\renewcommand{\theequation}{1.\arabic{equation}} \setcounter{equation}{0}

The discovery of gravitational waves (GWs) by the LIGO and Virgo collaborations in 2015 opened a new observational window on the universe and fundamentally transformed modern astronomy \cite{LIGOScientific:2016aoc, LIGOScientific:2016vbw}. Since then, GW astronomy has become an indispensable tool for investigating compact objects and strong-field gravity, offering direct access to regions where general relativity (GR) is tested in its most extreme form, such as binary black hole and binary neutron star mergers.

Besides these highly dynamical events, the motion of test particles around black holes provides an equally valuable arena for exploring relativistic dynamics. In particular, periodic orbits play a central role in organizing the phase space of bound motion. They offer a compact description of orbital structure and encode key information about stability, multi-scale dynamics, and interactions between black holes and surrounding matter \cite{Levin:2008mq, Levin:2009sk, Misra:2010pu, Babar:2017gsg}. More generally, generic orbits around black holes can often be viewed as perturbations of periodic trajectories \cite{Levin:2008mq}. For this reason, periodic motion and the radiation it generates are especially relevant in future GW observations.

A prominent example is the extreme mass ratio inspiral (EMRI), in which a stellar-mass black hole or neutron star gradually spirals into a supermassive black hole (SMBH). EMRIs are expected to be among the key targets of next-generation space-based detectors, including Taiji \cite{Hu:2017mde}, TianQin \cite{TianQin:2015yph, Gong:2021gvw}, and LISA \cite{Danzmann:1997hm, Schutz:1999xj, Gair:2004iv, LISA:2017pwj, Maselli:2021men}. Because the inspiral evolves slowly and the dissipated energy remains a tiny fraction of the total mass, the orbital motion of the smaller body can often be approximated by a succession of periodic segments over astrophysically long timescales. Consequently, the GW signal from an EMRI carries detailed information about both the orbital dynamics and the geometry of the central spacetime \cite{Bian:2025ifp}.

To trace the particle motion in such black hole geometries, one can implement the topological indexing scheme developed by Levin and Perez-Giz \cite{Levin:2008mq}. This framework assigns three characteristic integers, the number of radial oscillations (zooms $z$), relativistic revolutions (whirls $w$), and vertices $v$, to classify all closed coordinate trajectories.
This framework has been extensively applied to a broad range of spacetimes, including Schwarzschild and Kerr black holes \cite{Levin:2008ci, Levin:2009sk, Bambhaniya:2020zno, Rana:2019bsn}, charged black holes \cite{Misra:2010pu}, naked singularities \cite{Babar:2017gsg}, Kerr-Sen black holes \cite{Liu:2018vea}, and Horndeski hairy black holes \cite{Lin:2023rmo}. Related investigations in other backgrounds see Refs.~\cite{Wang:2022tfo, Lin:2023eyd, Haroon:2025rzx, Habibina:2022ztd, Zhang:2022psr, Lin:2022wda, Gao:2021arw, Lin:2021noq, Deng:2020yfm, Tu:2023xab, Zhou:2020zys, Azreg-Ainou:2020bfl, Wei:2019zdf, Pugliese:2013xfa, Zhang:2022zox, Healy:2009zm, Wang:2025wob, Alloqulov:2025bxh, Yao:2023ziq, Lin:2022llz, Chan:2025ocy, Wei:2025qlh, Sharipov:2025yfw} and references therein.

The associated GW emission from periodic motion has also attracted sustained attention, since it can preserve clear imprints of the underlying orbital parameters. A variety of waveform studies have been carried out in different black hole backgrounds; see, for instance, Refs.~\cite{Tu:2023xab, Yang:2024lmj, Shabbir:2025kqh, Junior:2024tmi, Zhao:2024exh, Jiang:2024cpe, Yang:2024cnd, Meng:2024cnq, Li:2024tld, QiQi:2024dwc, Haroon:2025rzx, Wang:2025hla, Lu:2025cxx, Zare:2025aek, Gong:2025mne, Li:2025sfe, Choudhury:2025qsh, Chen:2025aqh, Deng:2025wzz, Li:2025eln, Zahra:2025tdo, Ahmed:2025azu, Zhang:2025wni, Alloqulov:2025dqi, Lu:2025xlp, Zi:2023omh, Kumar:2024our, Kumar:2025njz, Kumar:2025jsi} and references therein.

On the other hand, although GR provides an excellent description of gravity, it is widely expected to be incomplete in regimes where quantum effects become important. In particular, singularities and the microscopic structure of spacetime remain unresolved within the classical theory \cite{Ashtekar:2021kfp}. This has motivated the study of effective black hole models incorporating quantum corrections. Among the approaches developed in this direction, the Hamiltonian-constraint framework \cite{Ashtekar:2004eh} plays a key role in canonical quantization and in the construction of quantum-modified geometries. A natural question is whether such corrections can leave observationally accessible signatures in astrophysical systems and thereby be constrained by current or future measurements.

Motivated by this question, a number of observational and phenomenological consequences of quantum-corrected black hole spacetimes have already been explored \cite{Konoplya:2024lch, Shu:2024tut, Skvortsova:2024msa, Liu:2024soc, Liu:2024wal, Jiang:2024cpe, Du:2024ujg, Ban:2024qsa, Wang:2024iwt, Uktamov:2024ckf, Nozari:2025veb, Zi:2024jla, Wang:2025fmz, Battista:2023iyu}. In parallel, the direct detection of GWs and the Event Horizon Telescope (EHT) observations of the SMBHs M87* \cite{EventHorizonTelescope:2019dse} and Sgr A* \cite{EventHorizonTelescope:2022wkp} have significantly expanded our ability to test gravity in the strong-field regime. These developments provide strong motivation to investigate whether quantum modifications of black hole spacetimes can imprint detectable signatures on orbital dynamics and GW emission.

In this work, we study GW radiation from periodic orbits of a test particle around a quantum-corrected Schwarzschild black hole. Our aim is to determine how the quantum deformation modifies the structure of bound motion and affects the resulting waveform. To this end, we adopt an effective spacetime inspired by loop quantum gravity \cite{Ashtekar:2021kfp,Ashtekar:2004eh}, analyze the deviations from the classical Schwarzschild dynamics, and compute the associated waveforms within the numerical kludge approximation \cite{Gair:2004iv}. We find that the quantum correction can produce noticeable changes in the phase evolution, waveform amplitude, and spectral content of the emitted radiation. These effects suggest that future space-based detectors such as LISA, Taiji, and TianQin may offer a promising avenue for probing quantum-gravity effects in the strong-field regime \cite{Hu:2017mde,TianQin:2015yph,LISA:2017pwj}.

This paper is organised as follows. In Sec.~\ref{section2}, we briefly review the quantum-corrected Schwarzschild spacetime. Sec.~\ref{section3} presents the dynamics of massive test particles in this geometry, with particular emphasis on periodic bound orbits. In Sec.~\ref{section4}, we compute the gravitational-wave emission from these trajectories and discuss the corresponding characteristic strain and Fourier spectra. Sec.~\ref{section5} concludes with a summary of the main results.

\section{Quantum corrected Schwarzschild black hole}\label{section2}
\renewcommand{\theequation}{2.\arabic{equation}} \setcounter{equation}{0}

In this section, we give a brief review of static, spherically symmetric black hole solutions in quantum gravity. From this perspective, Loop Quantum Gravity (LQG) emerges as one of the most promising quantum gravity proposals, aiming to regularise gravity \cite{Rovelli:2011eq, Ashtekar:2021kfp, Perez:2017cmj}. LQG is a non-perturbative theory for quantising the spacetime structure; however, it does not provide a complete quantum description of the removal of the singularity throughout spacetime. In this sense, one can investigate the effects of LQG in low-energy regimes through effective models, i.e., models that include corrections arising from quantum effects. Recently, in \cite{Alonso-Bardaji:2021yls, Alonso-Bardaji:2022ear}, the authors obtained a solution
of a regular, geodesically complete black hole, resulting from anomaly-free quantum corrections. Some aspects of this model have already been investigated, such as those linked to quasi-normal modes \cite{Fu:2023drp, Moreira:2023cxy, Bolokhov:2023bwm, Gingrich:2024tuf, Yang:2024ofe}, gravitational lensing \cite{Zhao:2024elr, Dong:2024alq, Soares:2024rhp, Liu:2024wal, Ahmed:2024fye, Shi:2024bpm}, and horizon area \cite{Sobrinho:2022zrp} \footnote{The phenomenological implications of quantum effects arising from LQG with different quantized approaches have also been extensively explored in refs.~\cite{Liu:2020ola, Zhu:2020tcf, Yan:2022fkr, Yan:2023vdg, Liu:2023vfh, Jiang:2023img, Jiang:2024vgn, Wang:2024iwt, Yang:2024cnd, Xamidov:2024xpc, Xamidov:2025oqx, Fatima:2025sdp} by several authors of this paper.}  In this work, we focus on gravitational radiation from periodic orbits around the quantum-corrected black hole. We begin with the line element \cite{Alonso-Bardaji:2021yls, Alonso-Bardaji:2022ear}
\begin{align}\label{metric1}
ds^2=-A(r)\,dt^2+\frac{r}{r-q_c} \frac{1}{A(r)}\,dr^2+r^2 d\Omega^2,
\end{align}
with
\begin{equation}\label{fr}
    A(r)=1-\frac{2M}{r},
\end{equation}
where $M$ denotes the black hole mass and $q_c$ is the holonomy-correction parameter, such that $q_c<2M$. It is a static, spherically symmetric, and asymptotically flat spacetime. The spacetime (\ref{metric1}) has a wormhole-like structure, with a minimal spacelike hypersurface separating the trapped regular black hole interior from the anti-trapped other region \cite{Soares:2023uup}. However, we consider light rays that do not cross the event horizon, characterised by $r = r_h = 2M$, i.e., we address regions with $r > r_h$. Though analytically, there is no clear difference between the event horizon of a black hole defined in ~(\ref{metric1}), and the standard Schwarzschild black hole, but there is a $g^{rr}$ term that is significantly different in both spacetimes. This small difference arises from quantum corrections and has a significant impact on particle motion. This difference motivates us to analyse the black hole solution with some quantum corrections. However, in the absence of a well-established theory of quantum gravity, these quantum-corrected models are constructed within the framework of semiclassical gravity. The quantum-correction parameter defined in (\ref{metric1}) is given by \cite{Soares:2023uup}: 
\begin{equation}
q_c=2M\frac{b^2}{1+b^2},
\end{equation}
 where $b$ is a dimensionless constant that is related to the fiducial length of the holonomies. It is worthwhile to mention that $q_c$ defines a minimal spacelike hypersurface, whose area is $4 \pi q_c^2$, separating the trapped black hole interior from the anti-trapped white hole region. The Komar, Arnowitt–Deser–Misner, and Misner-Sharp masses of the holonomy-corrected Schwarzschild black hole are given by \cite{Alonso-Bardaji:2022ear}
\begin{eqnarray}\label{mk-mass}
    M_K=M \sqrt{1-\frac{q_c}{r}}, \nonumber\\  M_{ADM}=M+\frac{q_c}{2}, \nonumber\\
    M_{MS}=M+\frac{q_c}{2}-\frac{M q_c}{r}.
\end{eqnarray}

\begin{figure*}
\includegraphics[width=8.2cm]{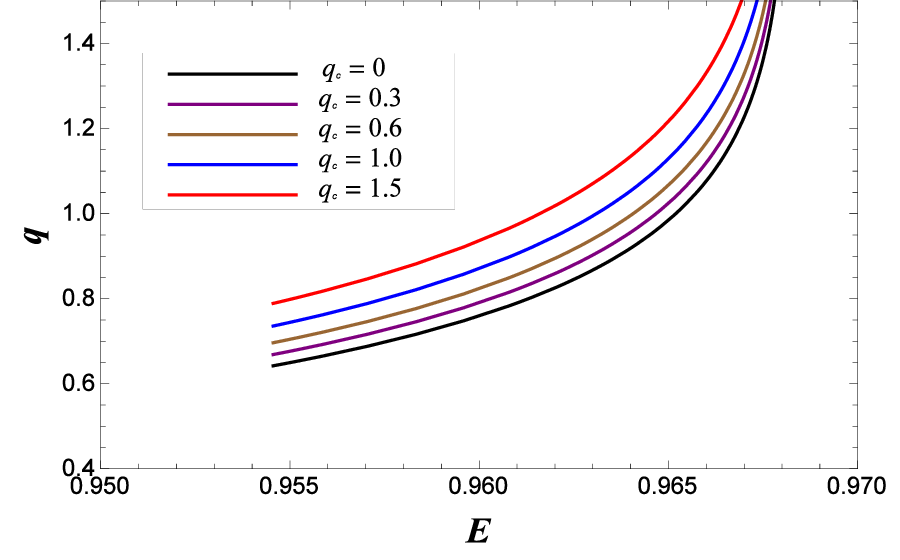}
\includegraphics[width=8.2cm]{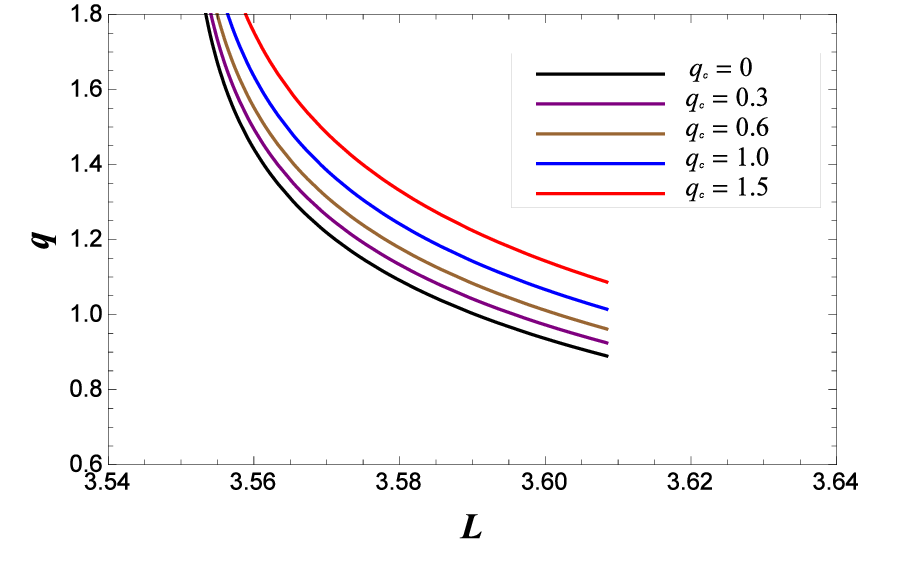}
\caption{The figure displays $q$ as a function of energy (left) and orbital angular momentum (right) across various values of $q_c$. The parameter settings are $L=3.73$ for the left panel and $E=0.95$ for the right panel. }
\label{qq1} 
\end{figure*}

In the limiting case $q_c \to 0$ or $b \to 0$, the spacetime is reverted to the standard Schwarzschild black hole. It should be noted that the presence of the minimal space-like hypersurface distinguishes the metric (\ref{metric1}) from the standard Schwarzschild black hole; otherwise, both form the event horizon at $r=2M$. Due to this, we can not differentiate the holonomy-corrected black hole from the Schwarzschild black hole, which motivates us to study the radial motion of the particle around the holonomy-corrected black hole, through which we can detect the difference through quantum corrections in the $g_{rr}$ term.

\begin{figure*}
\begin{tabular}{c c c}
\includegraphics[scale=0.55]{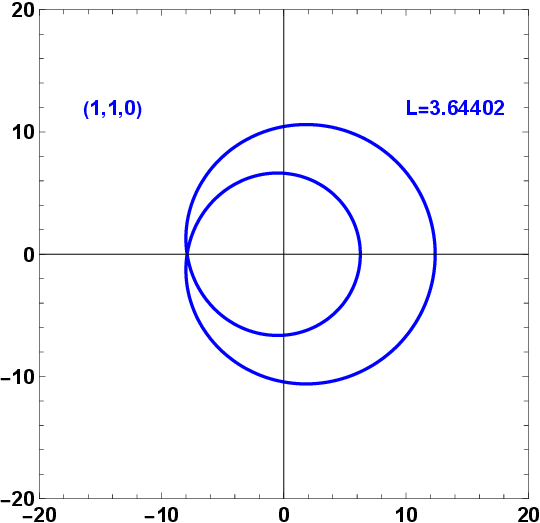}&
\includegraphics[scale=0.55]{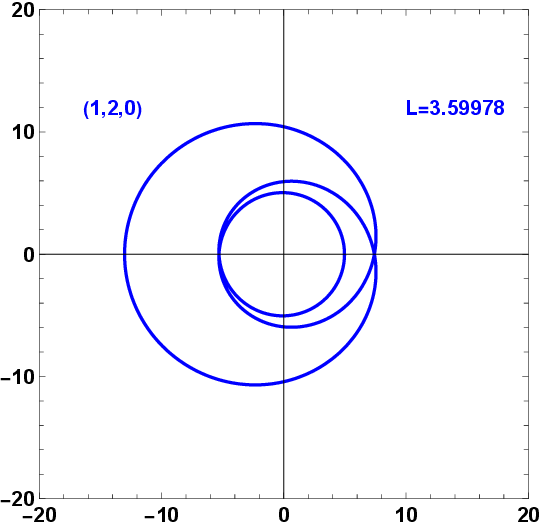}&
\includegraphics[scale=0.55]{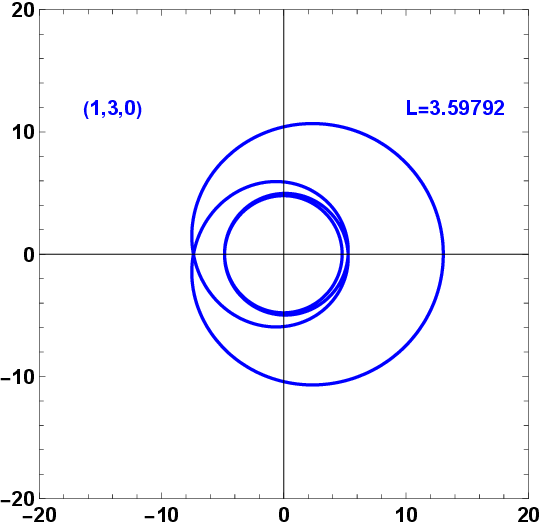}\\
\includegraphics[scale=0.55]{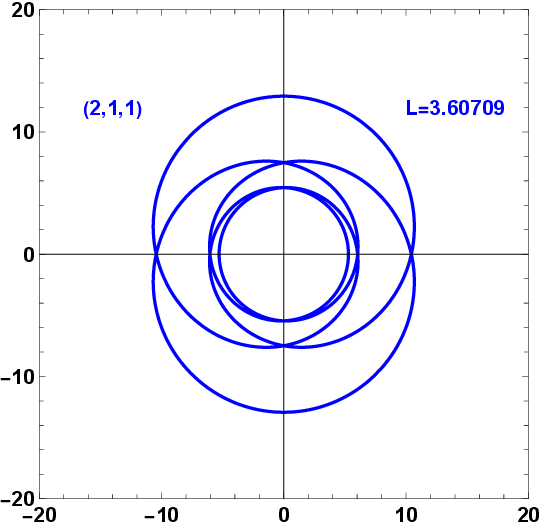}&
\includegraphics[scale=0.55]{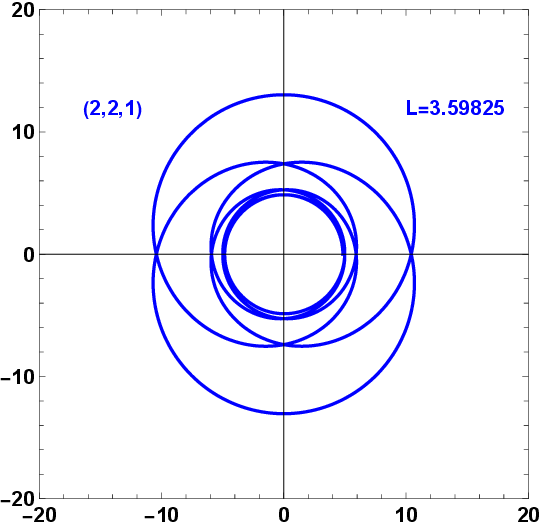}&
\includegraphics[scale=0.55]{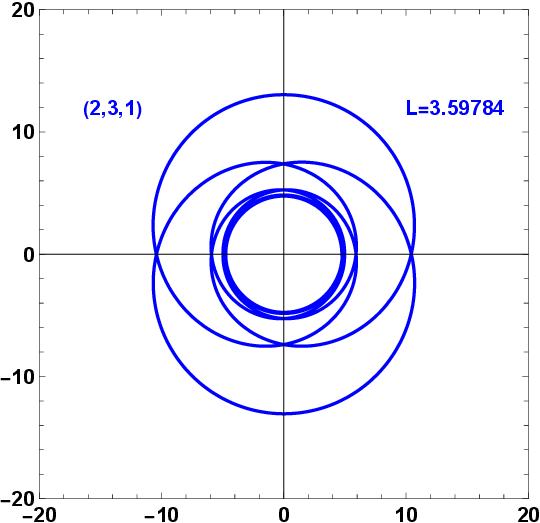}\\
\includegraphics[scale=0.55]{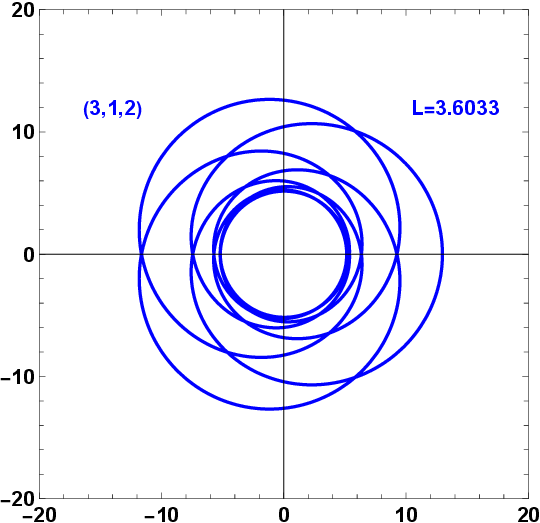}&
\includegraphics[scale=0.55]{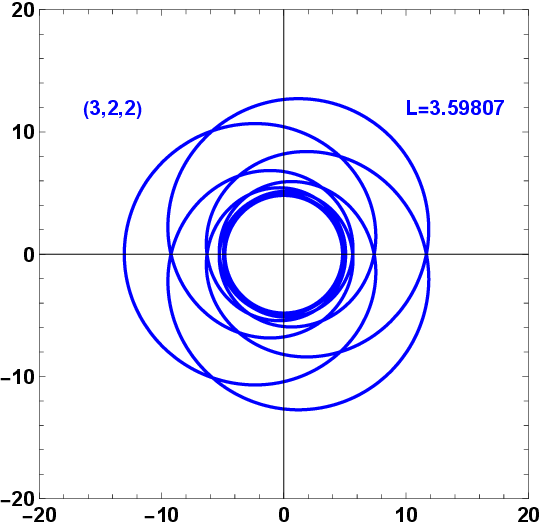}&
\includegraphics[scale=0.55]{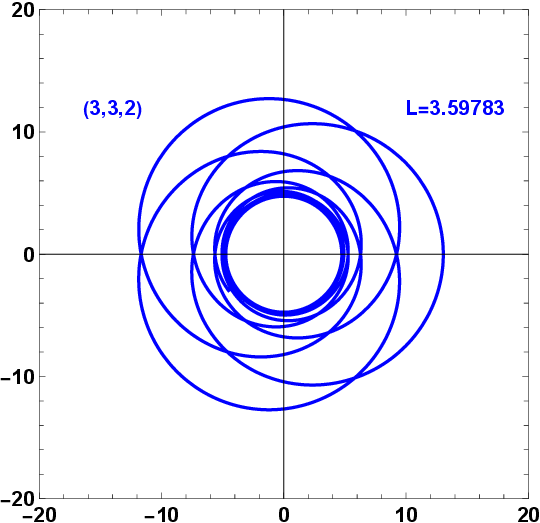}\\
\end{tabular}	
\caption{ Topological configurations of periodic orbits around a quantum-corrected black hole at fixed energy $E = 0.95$. Each profile highlights how varying sets of the topological numbers $(z, w, v)$ dictate the spatial complexity and geometric patterns of the bound state.}
\label{per-orb1} 
\end{figure*}

\section{Periodic orbits}\label{section3}
\renewcommand{\theequation}{3.\arabic{equation}} \setcounter{equation}{0}

In this section, we examine periodic timelike orbits in a quantum-corrected black hole spacetime. Since the structure of bound motion in strong gravitational fields is most naturally described through periodic trajectories, we begin with a brief review of geodesic motion \cite{Levin:2008yp}. Consider a test particle moving in a generic black hole background. Its dynamics are governed by the Lagrangian
\bqn
{\cal L} = \frac{1}{2} g_{\mu
u} \frac{{\rm d}x^\mu}{{\rm d}\lambda} \frac{{\rm d}x^
u}{{\rm d}\lambda},
\eqn
where \(\tau\) denotes the proper time for a timelike worldline and serves as the affine parameter. For null motion, \({\cal L}=0\), whereas for massive particles \({\cal L}<0\).

The canonical momentum associated with the coordinate \(x^\mu\) is
\bqn
p_\mu = \frac{\partial {\cal L}}{\partial \dot{x}^\mu} = g_{\mu
u}\dot{x}^
u.
\eqn
For a stationary and axisymmetric geometry, this yields the conserved quantities
\bqn
p_t &=& g_{tt}\dot{t}=-E,\\
p_\phi &=& g_{\phi\phi}\dot{\phi}=L,\\
p_r &=& g_{rr}\dot{r},\\
p_\theta &=& g_{\theta\theta}\dot{\theta},
\eqn
where \(E\) and \(L\) are the specific energy and angular momentum of the particle, respectively. Here and in the following, a dot denotes differentiation with respect to the affine parameter \(\lambda\).

Using these constants of motion, the temporal and azimuthal equations become
\bqn\lb{dot1}
\dot{t} = \frac{E}{A(r)},\\
\lb{dot2}
\dot{\phi} = \frac{L}{r^2\sin^2\theta}.
\eqn
For timelike geodesics, the normalization condition
\begin{equation}
g_{\mu
u}\dot{x}^\mu\dot{x}^
u=-1
\end{equation}
must also be satisfied. Substituting Eqs.~(\ref{dot1}) and (\ref{dot2}) into this relation gives
\bqn
g_{rr}\dot{r}^2 + g_{\theta\theta}\dot{\theta}^2
&=& -1 - g_{tt}\dot{t}^2 - g_{\phi\phi}\dot{\phi}^2  \nonumber \\
&=& -1 + \frac{E^2}{A(r)} - \frac{L^2}{r^2\sin^2\theta},
\eqn
which determines the radial and polar motion of the particle in the quantum-corrected background. This provides a convenient framework for analyzing zoom--whirl periodic trajectories and their potential observational signatures \cite{Levin:2008yp, Levin:2009sk, Misra:2010pu}.

We now restrict attention to equatorial motion, \(\theta=\pi/2\), with \(\dot{\theta}=0\). In this case, the radial equation reduces to
\bqn\lb{rdot}
\dot r^2 = (1-q_c/r)\bigl(E^2 - V_{\rm eff}(r)\bigr),
\eqn
where the effective potential is
\bqn \lb{Veff}
V_{\rm eff}(r)=\left(1+\frac{L^2}{r^2}\right)A(r).
\eqn
As \(r\to\infty\), one has \(V_{\rm eff}(r)\to 1\), consistent with asymptotic flatness. Accordingly, orbits with \(E>1\) are unbound and may escape to infinity, while \(E=1\) marks the threshold separating bound and unbound motion. Thus, \(E=1\) represents the largest energy compatible with bound trajectories. In principle, one can reconstruct the orbit by integrating Eqs.~(\ref{dot1}), (\ref{dot2}), and (\ref{Veff}) to obtain \(t(\tau)\), \(\phi(\tau)\), and \(r(\tau)\). However, because Eq.~(\ref{rdot}) contains a square root, its sign must be specified explicitly to distinguish inward from outward motion before numerical integration. For practical computations, it is therefore convenient to use the radial geodesic equation,
\bqn
\ddot{r}=\frac{A'(r)(1-q_c/r)+ A(r) q_c/r^2}{2(1-q_c/r)A(r)}\dot{r}^2 \nonumber\\ -\frac{A'(r)(1-q_c/r)E^2}{2A(r)}+\frac{A(r)(1-q_c/r)L^2}{r^3},
\eqn
which is well suited to numerical integration. This form also facilitates the study of circular-orbit stability and the transition to periodic or zoom-whirl motion in the strong-field regime \cite{Levin:2008yp, Levin:2009sk, Misra:2010pu, Chandrasekhar:1985kt}.

\begin{figure*}
\includegraphics[width=5.7cm]{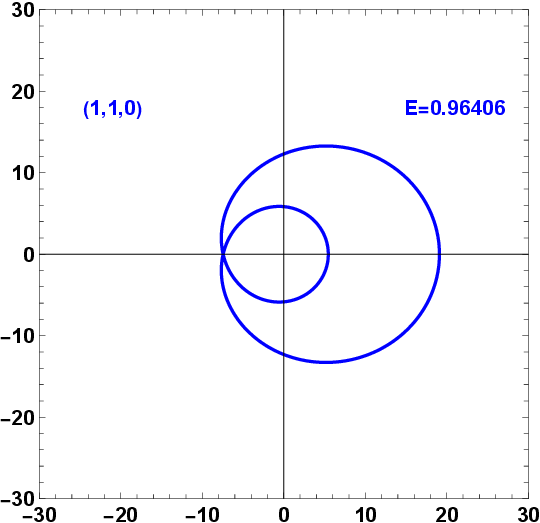}
\includegraphics[width=5.7cm]{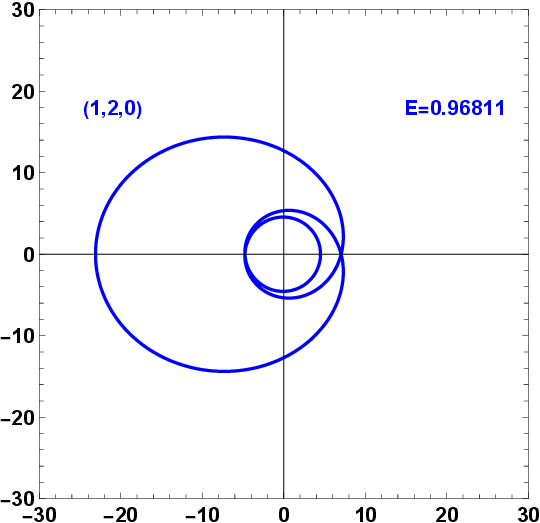}
\includegraphics[width=5.7cm]{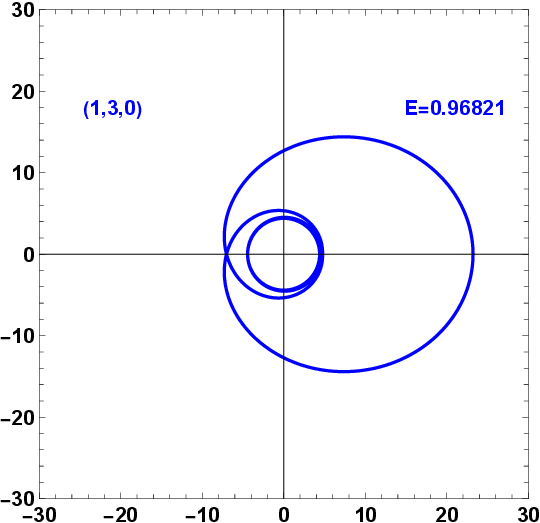}
\includegraphics[width=5.7cm]{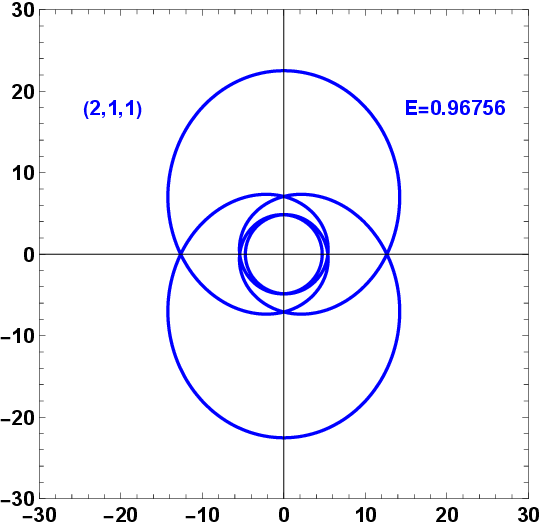}
\includegraphics[width=5.7cm]{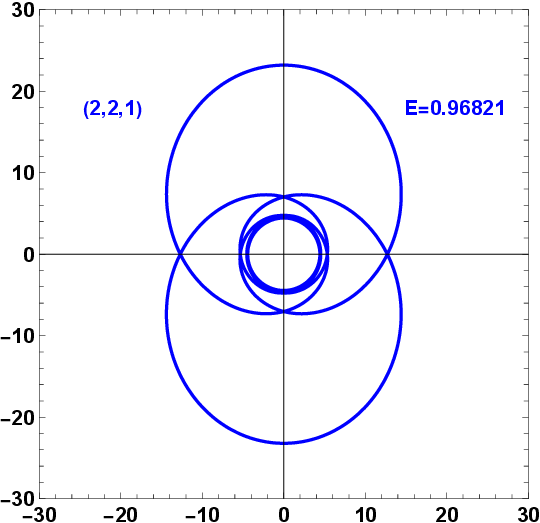}
\includegraphics[width=5.7cm]{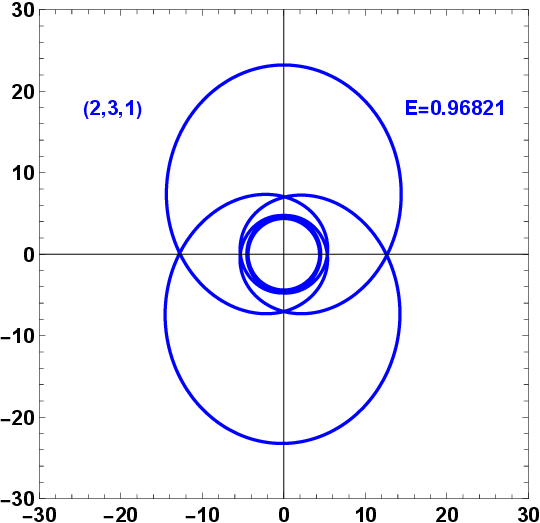}
\includegraphics[width=5.7cm]{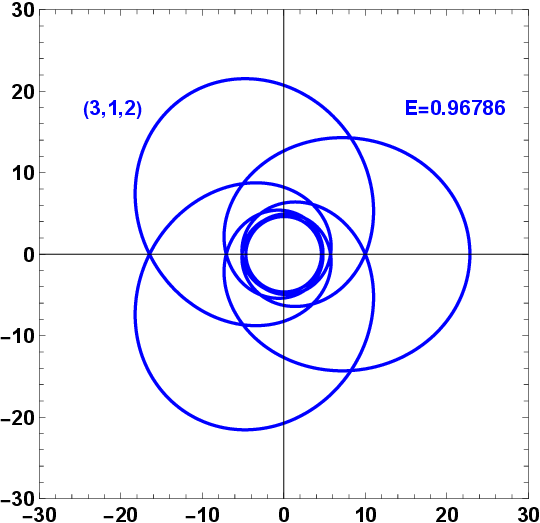}
\includegraphics[width=5.7cm]{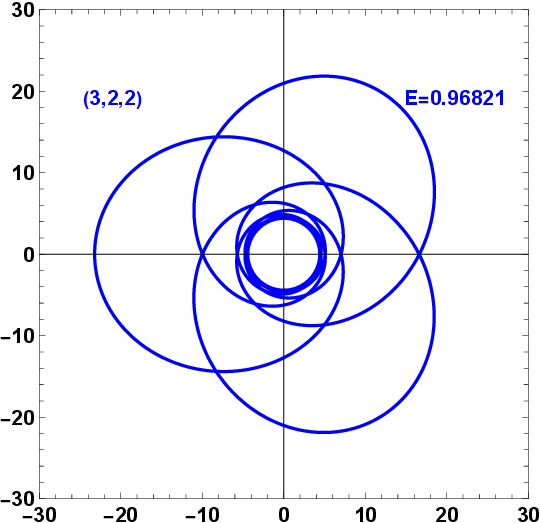}
\includegraphics[width=5.7cm]{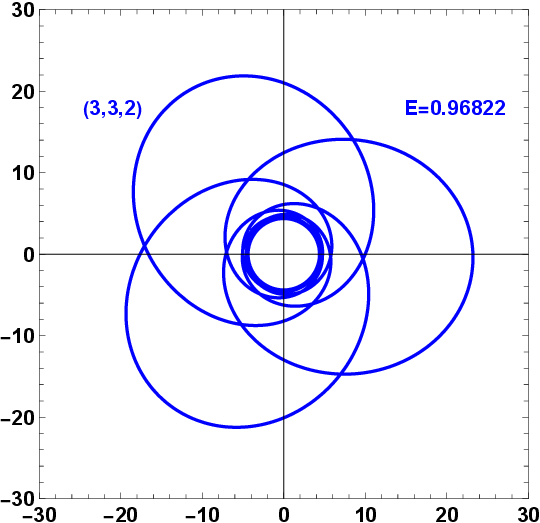}
\caption{Periodic orbits for various $(z, w, v)$ combinations around a quantum-corrected black hole. Here, we fixed the angular momentum at $L = 3.73$.}
\label{per-orb2} 
\end{figure*}

Once the equations of motion are integrated, periodic orbits can be obtained for suitable values of \(E\) and \(L\). Such orbits are bound and return to the same phase-space point after a finite period, although their geometric appearance can vary substantially with the orbital parameters. To classify them systematically, we adopt the scheme of Levin and Perez-Giz \cite{Levin:2008mq}. In this classification, each periodic orbit is labeled by a triplet \((z,w,v)\), where \(z\) counts the number of zooms, \(w\) the number of whirls, and \(v\) the vertex number. The integer \(v\) specifies the orientation in which the orbit visits successive vertices. To avoid degeneracy, \(z\) and \(v\) are chosen to be coprime \cite{Levin:2008mq}. The quantity
\bqn
q \equiv \frac{\omega_\phi}{\omega_r}-1 = w + \frac{v}{z}
\eqn
provides a compact measure of the deviation from a simple closed ellipse and encodes the ordering of the orbital segments. Since \(\omega_\phi/\omega_r = \Delta\phi/(2\pi)\), with \(\Delta\phi = \oint  {\rm d} \phi\), the value of \(q\) can be computed directly from the geodesic equations. For the quantum-corrected black hole spacetime considered here, one obtains
\bqn
q &=& \frac{1}{\pi} \int_{r_1}^{r_2} \frac{\dot \phi}{\dot r} {\rm d} r -1 \nonumber\\
&=& \frac{1}{\pi} \int_{r_2}^{r_1} \frac{L}{r^2\sqrt{(1-q_c/r)(E^2- V_{\rm eff}(r))}} {\rm d}r-1, \nonumber \\
\eqn
where \(r_1\) and \(r_2\) are the turning points of the radial motion.

Fig.~\ref{qq1} shows the dependence of \(q\) on \(E\) and \(L\). We find that, as the quantum correction parameter \(q_c\) increases, the allowed energy \(E\) increases monotonically, while the corresponding angular momentum \(L\) decreases. The admissible region in the \((E,L)\) plane is restricted, and periodic solutions exist only within this domain. These values determine the periodic orbits displayed in Figs.~\ref{per-orb1} and \ref{per-orb2} for different choices of \((z,w,v)\). As expected, larger \(z\) corresponds to more pronounced multi-petal structure and a more intricate zoom--whirl pattern. In Fig.~\ref{per-orb1}, the energy is held fixed, whereas in Fig.~\ref{per-orb2}, the angular momentum is fixed. For simplicity, we set \(M=1\) in these plots.

\section{Gravitational Radiation in quantum-corrected geometry}\label{section4}
\renewcommand{\theequation}{4.\arabic{equation}} \setcounter{equation}{0}

Next, we turn to the gravitational radiation produced by a test particle on periodic orbits around a quantum-corrected SMBH. Extreme mass ratio inspirals (EMRIs), in which a stellar-mass compact object slowly moves around a supermassive black hole, are among the most promising sources for space-based GW observatories such as LISA, Taiji, and TianQin \cite{LISA:2017pwj, Gong:2021gvw, Hu:2017mde}. The waveform emitted by such systems carries detailed information about both the orbital dynamics and the geometry of the central spacetime. In this sense, EMRIs may provide a sensitive probe of possible quantum modifications to black hole backgrounds and are therefore of considerable interest for future observations.

\begin{figure*}
\includegraphics[width=8cm]{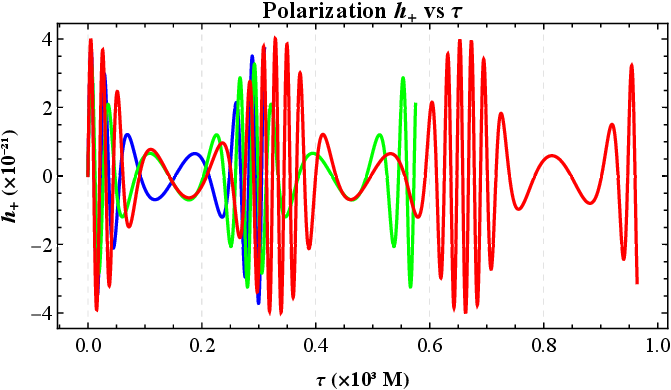}
\includegraphics[width=8cm]{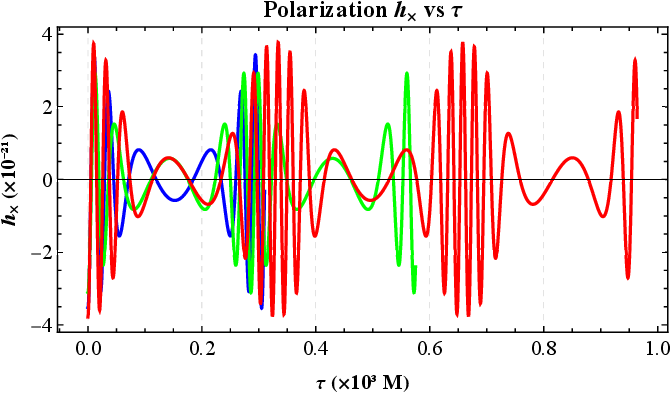}
\caption{Simulated GW signatures ($h_+$ and $h_\times$ modes) showing distinct zoom–whirl phases on the amplitude and frequency profiles. The background is a quantum-corrected black hole ($M = 10^{6} M_{\odot}$, $q_c = 0.6$), and the radiating test mass is $m = 10M_{\odot}$ with $E = 0.95$. Three separate periodic configurations are color-coded: $(1,2,0)$ in blue, $(2,1,1)$ in green, and $(3,2,2)$ in red.}
\label{gwpolar1} 
\end{figure*}

Our simulation of EMRI waveforms relies on the numerical kludge (NK) prescription \cite{Babak:2006uv}, operating under the standard adiabatic approximation. This hybrid model decouples the waveform generation into two core procedures: (i) integrating the exact equations of motion derived from the LQG-corrected background to map the particle trajectory, and (ii) projecting this coordinate path into flat spacetime to evaluate the gravitational radiation via the conventional quadrupole formula.

For a gravitational waveform \(h_{ij}\) and a symmetric trace-free (STF) mass quadrupole moment \(I_{ij}\), the quadrupole formula reads
\begin{equation}
h_{ij}=\frac{1}{A}\ddot{I}_{ij},
\end{equation}
where \(A=c^4 D_L/(2G)\), and we set \(G=c=1\), with \(D_L\) denoting the luminosity distance to the source. Solving the geodesic equations numerically yields the worldline \(Z^i(t)\) of the small body in the curved spacetime of the SMBH, which is then used to construct the emitted GW signal. For a particle of mass \(m\) following the trajectory \(Z^i(t)\), the mass quadrupole moment is given by \cite{Thorne:1980ru}
\begin{equation}\label{lvalue}
I^{ij}=m\int {\rm d}^3x\, x^i x^j\, \delta^3(x^i - Z^i(t)).
\end{equation}

The choice of coordinates plays an important role in both the construction and interpretation of GW signals. Although the geodesic equations are most naturally integrated in spherical coordinates \((r,\theta,\phi)\), the waveform is more conveniently expressed in a Cartesian source frame \((x,y,z)\), which is also suitable for comparison with detector data. The corresponding coordinate transformation is given by \cite{Babak:2006uv}
\begin{equation}\label{4.3}
x = r \sin\theta \cos\phi, \quad 
y = r \sin\theta \sin\phi, \quad 
z = r \cos\theta.
\end{equation}
This mapping allows the particle trajectory to be projected onto a Cartesian grid, which is needed when evaluating the source multipole moments. In the numerical kludge framework, the metric perturbation \(h_{ij}\), representing the emitted GW, is obtained from the second time derivative of the mass quadrupole moment \(I_{ij}\), or equivalently,
\begin{equation}\label{4.4}
h_{ij} = \frac{m}{A}\left(a_i x_j + a_j x_i + 2 v_i v_j\right),
\end{equation}
where \(v_i\) and \(a_i\) denote the velocity and acceleration of the small body, respectively, and \(A=c^4 D_L/(2G)\) with \(G=c=1\). This is the standard form used in numerical kludge waveform calculations \cite{Barack:2003fp, Gair:2004iv, Babak:2006uv, Hughes:2000ssa} and provides a practical and physically consistent approximation for EMRI signals.

\begin{figure*}
\includegraphics[width=8.2cm]{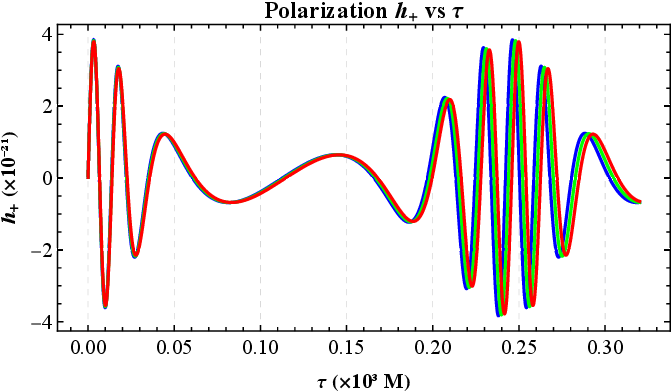}
\includegraphics[width=8.2cm]{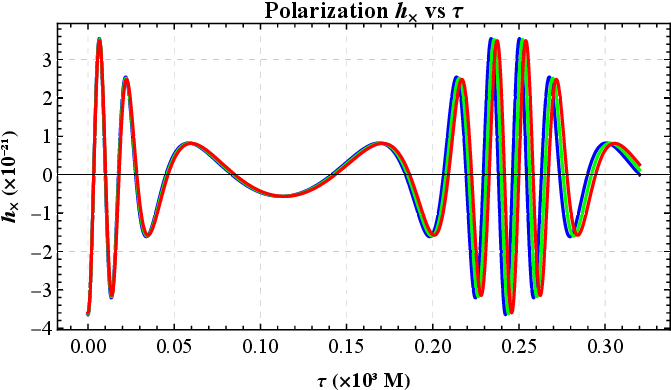}
\caption{
Simulated GW signatures ($h_+$ and $h_\times$ modes) of the quantum corrected  SMBH with $M = 10^{6} M_{\odot}$ and the radiating test mass  $m = 10M_{\odot}$ traveling in a $(1,2,0)$ periodic orbit with $E = 0.95$. Three separate parameter configurations are color-coded: $q_c = 0$ (Schwarzschild case) in blue, $0.2$ in green, and $0.4$ in red.
} 
\label{gwpolar2} 
\end{figure*}

To connect the source description to the detector response, it is useful to introduce a detector-oriented Cartesian frame \((X,Y,Z)\), centered on the black hole and related to the source frame \((x,y,z)\) by the inclination angle \(\iota\) and the longitude of pericentre \(\zeta\) \cite{Babak:2006uv, Barack:2003fp, Gair:2004iv}. This rotation enables one to project the waveform into the detector frame and extract the observable GW polarizations. The corresponding unit vectors are
\begin{eqnarray}
\hat{e}_X &=& (\cos\zeta, -\sin\zeta, 0),\\
\hat{e}_Y &=& (\sin\iota \sin\zeta, \cos\iota \cos\zeta, -\sin\iota),\\
\hat{e}_Z &=& (\sin\iota \sin\zeta, -\sin\iota \cos\zeta, \cos\iota),
\end{eqnarray}
and the two GW polarizations are obtained by projecting \(h_{ij}\) in Eq.~(\ref{4.4}) onto this basis:
\begin{eqnarray}\label{4.5}
h_+&=\frac{1}{2}\big(e_X^i e_X^j-e_Y^ie_Y^j\big)h_{ij},\\
h_{\times}&=\frac{1}{2}\big(e_X^i e_Y^j-e_Y^ie_X^j\big)h_{ij}.
\end{eqnarray}
Equivalently, the polarizations may be written in terms of the detector-frame components \(h_{\zeta\zeta}\), \(h_{\iota\iota}\), and \(h_{\iota\zeta}\) as
\begin{eqnarray}\label{4.64}
h_+&=&\frac{1}{2}\big(h_{\zeta\zeta}-h_{\iota\iota}\big),\\\label{4.6}
h_{\times}&=&h_{\iota\zeta},
\end{eqnarray}
with
\begin{eqnarray}\label{4.7}
h_{\zeta\zeta}&=&h_{xx}\cos^2\zeta-h_{xy}\sin{2 \zeta}+h_{yy}\sin^2\zeta,\\
h_{\iota\iota}&=& \cos^2\iota\big[h_{xx}\sin^2\zeta + h_{xy}\sin 2 \zeta + h_{yy}\cos^2 \zeta\big] 
\nonumber \\
&& +h_{zz} \sin^2\iota - \sin{2 \iota}\big[h_{xz \sin\zeta}+h_{yz}\cos\zeta\big],\\
h_{\iota\zeta}&=& \frac{1}{2}\cos\iota\big[h_{xx} \sin {2 \zeta}+ 2 h_{xy}\cos{2 \zeta}- h_{yy}\sin{2 \zeta}\big]
\nonumber\\
&& +\sin\iota\big[h_{yz} \sin\zeta-h_{xz} \cos\zeta\big].
\end{eqnarray}

\begin{figure*}
\includegraphics[width=8cm]{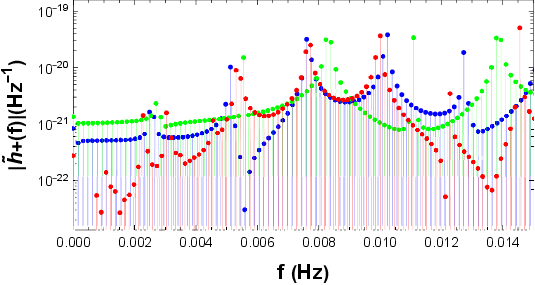}
\includegraphics[width=8cm]{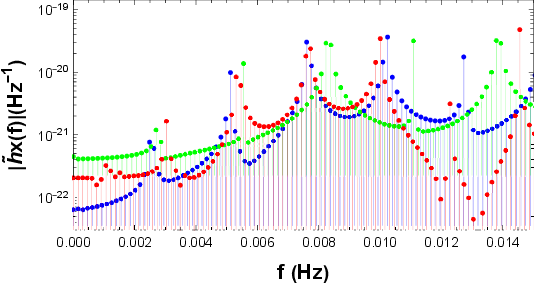}
\caption{
Fourier spectra $|\tilde{h}_{+,\times}(f)|$ of the radiation profiles from Fig.~\ref{gwpolar1} with loop-quantum parameter $q_c = 0.6$. The resonant peaks trace the characteristic frequency spectrum of the periodic motion, where the structural variations in the harmonics are governed by the specific triad $(z, w, v)$.
}
\label{freq-spect1} 
\end{figure*}

\begin{figure*}
\includegraphics[width=8cm]{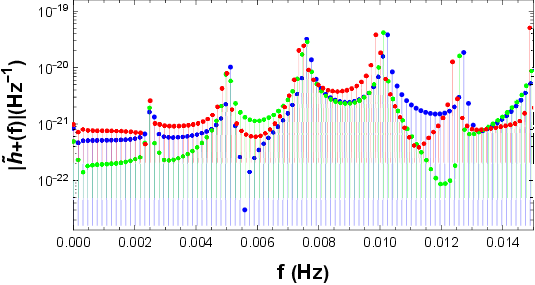} 
\includegraphics[width=8cm]{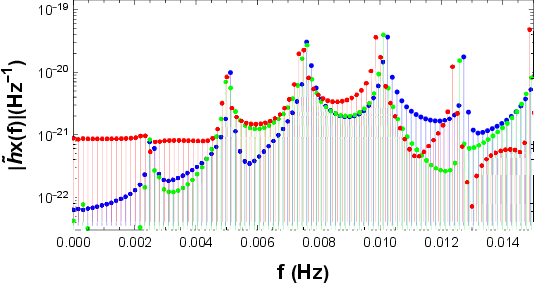}
\caption{Fourier spectra $|\tilde{h}_{+,\times}(f)|$ for the waveforms in Fig.~\ref{gwpolar2} with zoom-whirl value $(1,2,0))$. Increasing the quantum-correction parameter shifts the spectral peaks and enhances the high-frequency components, indicating stronger gravitational radiation and modified orbital dynamics.}
\label{freq-spect2} 
\end{figure*}

To investigate how the quantum-correction parameter affects the waveforms produced by different periodic orbits in an EMRI system, we consider a compact object with mass \(m=10\,M_\odot\) orbiting an SMBH of mass \(M=10^6\,M_\odot\). For simplicity, we set the inclination angle and the longitude of pericentre to \(\iota=\zeta=\pi/4\), and adopt a luminosity distance \(D_L=200\) Mpc when computing the GW polarizations.

The emitted signals clearly record the history of the eccentric trajectory, appearing as distinct, modulated wave packets. When the test particle recedes toward the apastron (the zoom segment), the gravitational flux diminishes, yielding low-amplitude oscillatory tails. Conversely, as the particle plunges into the high-curvature region near the horizon (the whirl segment), the radiation intensifies sharply into high-energy pulses.

In Fig.~\ref{gwpolar1}, we display the waveforms corresponding to \((z,w,v)=(1,2,0)\), \((2,1,1)\), and \((3,2,2)\). The results show a direct correspondence between the waveform morphology and the underlying orbital trajectory. Each periodic orbit generates a waveform with distinct zoom and whirl segments, mirroring the motion of the particle in the spacetime. Moreover, larger values of \(z\) lead to richer waveform substructures, reflecting the increased number of petals in the associated periodic orbit.

The quantum-correction parameter \(q_c\) also has a visible impact on the waveform generated by the orbiting particle. In Fig.~\ref{gwpolar2}, we show the waveforms for several values of \(q_c\) and compare them with the Schwarzschild case ($q_c=0$). Relative to the Schwarzschild case ($q_c=0$), the waveforms for $q_c=0.2$ and $q_c=0.4$ exhibit small but systematic deviations in both amplitude and phase, which grow monotonically with $q_c$ and signal a quantum-correction-induced dephasing, consistent with the spectral-peak shifts seen in Fig.~\ref{freq-spect2}. This indicates that the quantum deformation of the background spacetime modifies the orbital dynamics and leaves measurable imprints on the emitted radiation.

The GWs produced by a particle in periodic motion around a quantum-corrected SMBH may also be analyzed in the frequency domain through the spectra \(\lvert \tilde{h}_{+,\times}(f)\rvert\) and the characteristic strain \(h_c(f)\), defined as
\begin{eqnarray}\label{ch}
h_c(f)=2f\left(\vert \tilde{h}_+(f)\vert^2+\vert \tilde{h}_\times(f)\vert^2\right)^{1/2}.
\end{eqnarray}
We obtain these spectra using a discrete Fourier transform of the time-domain waveforms, which allows us to examine the frequency content of the signal in detail. In particular, the periodic orbital motion determines the structure of the emitted radiation in frequency space, as illustrated in Figs.~\ref{freq-spect1} and~\ref{freq-spect2}. The dominant frequencies lie mainly in the millihertz band, making them especially relevant for space-based detectors such as LISA, Taiji, and TianQin \cite{LISA:2017pwj, Barack:2003fp, Gair:2004iv, Babak:2006uv}, which are designed to probe low-frequency GW sources such as EMRIs. The characteristic spectra for different periodic orbits labeled by \((z,w,v)\) are shown in Fig.~\ref{freq-spect1}, while those for different values of \(q_c\) are presented in Fig.~\ref{freq-spect2}.

\begin{figure*}
\includegraphics[width=8.2cm]{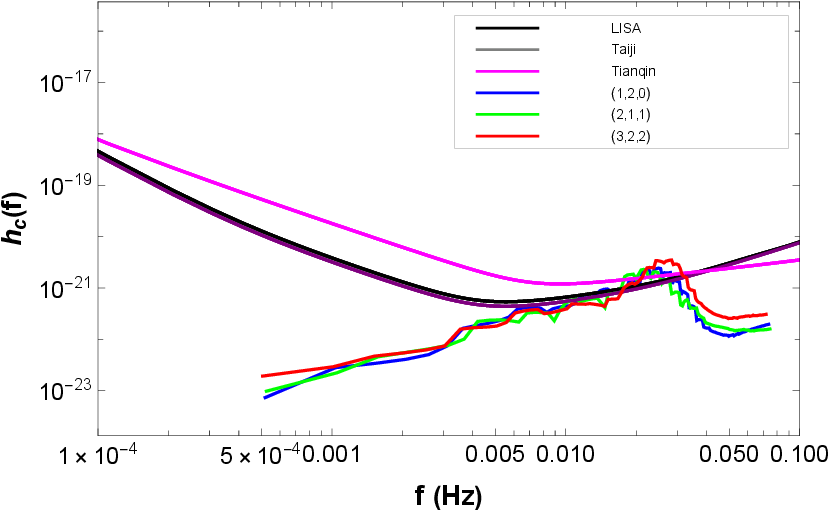} 
\includegraphics[width=8.2cm]{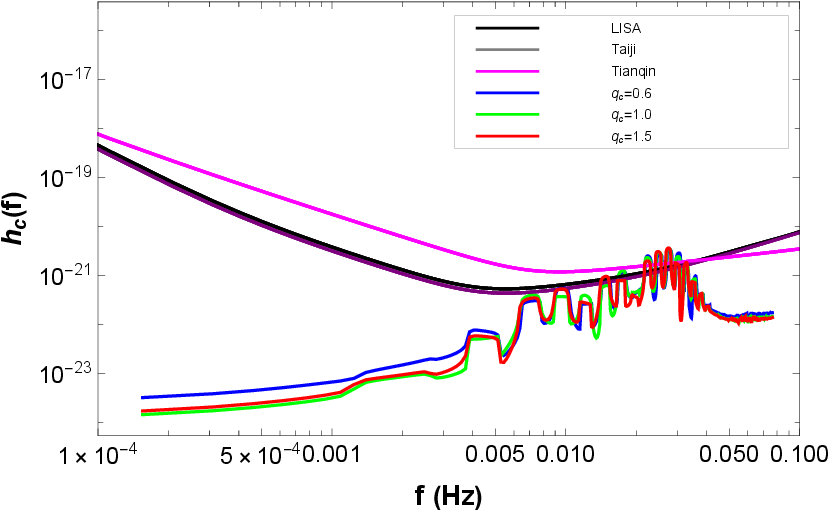}
\caption{Characteristics strain of gravitational waveforms of periodic orbits in Figs.~\ref{gwpolar1} (left) and \ref{gwpolar2} (right). The black, gray, and magenta curves correspond to the sensitivities of the LISA, Taiji, and Tianqin detectors. Portions of the spectra lie above the sensitivity bands, suggesting that these quantum-correction effects could be detectable in future space-based GW observations.}
\label{strain} 
\end{figure*}

To improve the visual presentation of the characteristic strain defined in Eq.~(\ref{ch}), we apply a smoothing procedure to the numerically computed \(h_c(f)\) by taking a running average over 30 frequency bins. A larger averaging window would suppress numerical fluctuations more effectively, but it could also wash out some of the finer spectral details. As illustrated in Fig.~\ref{strain}, the characteristic strain corresponding to different orbital configurations \((z,w,v)\) and values of the quantum-correction parameter \(q_c\) lies above the sensitivity curves of LISA, Taiji, and TianQin. This indicates that the associated GWs, which retain the zoom--whirl imprint of the underlying orbital motion, fall within the observable range of these planned space-based detectors \cite{LISA:2017pwj, Gair:2017ynp, Babak:2017tow}. A detection of such signals would offer a valuable opportunity to probe the near-horizon spacetime of SMBHs and to search for potential quantum-gravity effects through high-precision GW measurements.

\section{Discussions and Conclusions}\label{section5}

In this paper, we have modeled the EMRI system under a loop-quantum-corrected Schwarzschild background to evaluate how quantum gravity imprints on gravitational radiation. Our numerical results confirm that the holonomy parameter $q_c$ induces substantial alterations in both the orbital frequency and the temporal phase structure of the waveforms, making these quantum signatures potentially observable.

We focused on periodic orbits and their associated waveforms in a quantum-corrected spacetime. The geodesic equations were solved analytically in the background geometry, and the periodic trajectories were classified using the representation of Levin and Perez-Giz \cite{Levin:2008mq}. In this framework, each periodic orbit is characterized by the integers \((z,w,v)\), which provide a geometric description of the orbital structure.

We then studied the gravitational waveforms generated by periodic motion in the quantum-corrected geometry. These results suggest that waveform morphology may be used to distinguish quantum-corrected black holes from the Schwarzschild case. In our EMRI setup, a compact object of mass \(m=10\,M_\odot\) was assumed to orbit an SMBH of mass \(M=10^6\,M_\odot\). Using the numerical kludge approach, we computed the gravitational waveforms with the source placed at a luminosity distance of \(D_L=200\) Mpc, and with inclination angle \(\iota=\pi/4\) and longitude of pericenter \(\zeta=\pi/4\). The resulting signals exhibit a clear correspondence with the zoom--whirl behavior of the orbiting body: more complicated orbital structures, associated with larger zoom--whirl numbers, produce richer waveform substructures. We also found that the quantum-correction parameter \(q_c\) significantly affects the waveform. To assess detectability, we performed discrete Fourier transforms of the time-domain signals and extracted the frequency spectra. The dominant frequencies fall within the sensitivity band of planned space-based detectors. From the spectra, we also computed the characteristic strain and found that, for certain choices of \((z,w,v)\), the signal lies above the noise curves of LISA, Taiji, and TianQin. This indicates that periodic-orbit EMRIs in quantum-corrected spacetimes may be observable with next-generation GW missions, offering a promising avenue to probe SMBHs in the presence of quantum effects.

Finally, we emphasize that the present analysis is exploratory in character: the waveforms are computed in a non-dissipative framework (exact geodesics plus the quadrupole formula, without radiation reaction), and the values of the quantum-correction parameter $q_c$ adopted here are chosen for clarity of illustration rather than as physically motivated constraints. A quantitative assessment of the observation time required to distinguish quantum-corrected from Schwarzschild waveforms in realistic EMRIs would require a self-consistent treatment of radiation reaction together with physically motivated (likely smaller) values of $q_c$; we defer such an analysis to future work.

Several other generalizations of the current model can be explored in future investigations. A crucial next step is to incorporate radiation-reaction forces to track the self-force-driven inspiral trajectory in this quantum-corrected geometry. Furthermore, resolving the waveform beyond the quadrupole formula by extracting higher multipole contributions will enhance template accuracy. Most importantly, expanding this analysis to a rotating, loop-quantum-corrected (Kerr-like) black hole background represents a vital path toward realistic astrophysical applications.

\section*{Acknowledgements}
 
This work is supported by the National Natural Science Foundation of China under Grants No. 12275238, No. 12542053, and No. 11675143, the National Key Research and Development Program under Grant No. 2020YFC2201503, and the Zhejiang Provincial Natural Science Foundation of China under Grants No. LR21A050001 and No. LY20A050002, and the Fundamental Research Funds for the Provincial Universities of Zhejiang in China under Grant No. RF-A2019015.

\bibliography{main}

\end{document}